# Modelling by maps of two-frequency microwave ionization of hydrogen atoms

B. Kaulakys[1], D. Grauzhinis and G. Vilutis

*Institute of Theoretical Physics and Astronomy, A. Goštauto 12, 2600 Vilnius, Lithuania*

**Abstract**. -Mapping equations of motion of the highly exited classical atom in a monochromatic field are generalized for the two-frequency microwave field. Analysis of the obtained equations indicates to the weak sensitivity of the position of the recently observed ionization peak near the main resonance to the frequency and amplitude of the additional microwave field. In the high frequency region, however, the sensitivity of the enhanced ionization peaks on the additional field frequency is predicted.

At present studies of the highly excited atoms in an intense electromagnetic field attract a large amount of effort. The Rydberg atom in a monochromatic field is one of the simplest real quasiclassical systems with stochastic behavior which may be investigated both theoretically and experimentally [1]-[3]. Classical regular motion and the stochastic dynamics of the excited electron in the monochromatic field resulting in the diffusion-like excitation and ionization processes may be described by the map, called the 'Kepler map' [2]-[6]. This greatly facilitates numerical and analytical investigations of transition to the stochasticity and ionization process. It appears that, although the derivation of the Kepler map is based on the classical perturbation theory, the map is, nevertheless, suitable even for the low frequency field when at transition to the chaotic behavior the strength of the ionizing field is comparable with the Coulomb field [6].

The main objective of the present work is to derive and investigate the mapping equations of the motion for the highly excited atom in two- and multi-frequency fields. The urgency of this problem follows also from the experimental and theoretical investigations [7], where nonmonotonic behavior (peak) in the two-frequency ionization has been observed. The relatively simple mapping form of the equations of motion allows theoretical investigation and predictions of similar peculiarities for different parameters of the problem.

We start from the minimal coupling [6] of the electromagnetic field to the electron through the $\mathbf{A} \cdot \mathbf{P}$ interaction where $\mathbf{A}$ is the vector potential of the field and $\mathbf{P}$ is the generalized momentum of the electron. From the Hamiltonian of the hydrogen atom in a linearly polarized total external field $\mathbf{F}_T$ we can obtain an equation for the electron energy $E$ change due to the interaction with this field [4]-[6]

$$\dot{E} = -\dot{\mathbf{r}} \cdot \mathbf{F}_T. \tag{1}$$

Here $\dot{\mathbf{r}}$ is the electron velocity. The full field strength may be expressed through the components

as
$$\mathbf{F}_T = \sum_k \mathbf{F}_k \cos(\omega_k t + \vartheta_k) \qquad (2)$$

where $\mathbf{F}_k$, $\omega_k$ and $\vartheta_k$ are the amplitudes, frequencies and phases of the field components, respectively. Further for the simplicity we restrict ourselves to the two-frequency field parallel to the $x$-axis

$$F_T = F\cos(\omega t + \vartheta) + F_2 \cos(\omega_2 t + \vartheta_2) \qquad (3)$$

and consider widely used in theoretical analysis [2]-[8] the one-dimensional model which corresponds to the states very extended along the electric field direction. We remain the same usual [4]-[6], [8] notation for the parameters of first (referee) field, i.e. $\mathbf{F}$, $\omega$ and $\vartheta$ without the subscripts, while the second field parameters supply with subscripts 2.

To minimize the number of the free parameters it is convenient [5]-[6], [8] to introduce some scaled and relative quantities: the scaled energy $\varepsilon = -2E/\omega^{2/3}$, the relative field strengths $F_0 = Fn_0^4$ and $F_{2,0} = F_2 n_0^4$ (with $n_0 = (-2E_0)^{-1/2}$ being the initial principle quantum number of the hydrogen atom) and the relative field frequencies $s = \varepsilon^{-3/2} = \frac{\omega}{(-2E)^{3/2}}$ and $s_2 = s/\kappa = \frac{\omega_2}{(-2E)^{3/2}}$. Here $\kappa = \omega/\omega_2 = s/s_2$ is the ratio of two microwave frequencies while $s$ and $s_2$ are the ratios between the microwave frequencies, $\omega$ and $\omega_2$, and the Kepler orbital frequency $\Omega = (-2E)^{3/2}$, respectively. Then, using the parametric equations of the unperturbed motion of the electron

$$\begin{cases} x = \frac{1-\cos\xi}{-2E}, \\ t = \frac{\xi - \sin\xi}{(-2E)^{3/2}}, \end{cases} \qquad (4)$$

and integrating eq. (1) for the electron motion between two subsequent passages at the aphelion (where $\dot{x} = 0$ and there is no electron's energy change) we obtain the map for the scaled energy $\varepsilon$ and the field phase $\vartheta$ at the electron's passage of the perihelion, $x = 0$, moment (see [4]-[6] for analogy):

$$\begin{cases} \varepsilon_{j+1} = \varepsilon_j - \pi\varepsilon_0^2\Big[F_0 h(\varepsilon_{j+1})\sin\vartheta_j + \\ \qquad\qquad + F_{2,0}\kappa^{2/3} h(\kappa^{2/3}\varepsilon_{j+1})\sin\vartheta_{2,j}\Big], \\ \vartheta_{j+1} = \vartheta_j + 2\pi\varepsilon_{j+1}^{-3/2} - \pi\varepsilon_0^2\Big[F_0\eta(\varepsilon_{j+1})\cos\vartheta_j + \\ \qquad\qquad + F_{2,0}\kappa^{7/3}\eta(\kappa^{2/3}\varepsilon_{j+1})\cos\vartheta_{2,j}\Big]. \end{cases} \qquad (5)$$

Here

$$h(\varepsilon_{j+1}) = \frac{4}{\varepsilon_{j+1}} \mathbf{J}'_{s_{j+1}}(s_{j+1}), \quad \eta(\varepsilon_{j+1}) = \frac{dh(\varepsilon_{j+1})}{d\varepsilon_{j+1}}, \qquad (6)$$

$\vartheta_{2,j} = \vartheta_{2,0} - \vartheta_0/\kappa + \vartheta_j/\kappa$ and $\mathbf{J}'_s(z)$ is the derivative of the Anger function. Note that here $\varepsilon_0 = -2E_0/\omega^{2/3}$, $s_0 = \omega n_0^3 = \varepsilon_0^{-3/2}$ and $s_{2,0} = s_0/\kappa$ indicate the initial values of the scaled energy and relative field frequencies, respectively, while $\varepsilon_j$, $s_j$ and $s_{2,j} = s_j/\kappa$ correspond to the current values of the variables.

The first equation of the map (5) contains two terms corresponding to the change of the electron's scaled energy during the intrinsic motion period by the influence of two field components, respectively, while the second equation follows from the requirement of the area-preserving of the map. Note also that expression $\kappa^{2/3}\varepsilon_{j+1}(=-2E_{j+1}/\omega_2^{2/3} \equiv \varepsilon_{2,j+1})$ in eqs. (5) is in fact the electron energy scaled according to the frequency $\omega_2$ of the second field component.

We have, therefore, derived the two-dimensional map for the dynamics of the hydrogen atom in the two-frequency field similar to that for the monochromatic field [4]-[6]. Taking into account condition that phases of the field components are all time interrelated we can easily generalize map

(5) to the multi-frequency fields as well; simply adding terms to the right-hand side of eqs. (5) corresponding to the additional field components.

The analysis of chaotic dynamics and ionization process described by map (5)-(6) is, therefore, quite similar to that of the simple Kepler map [4]-[6]. So, we can easily analyze dependencies of the dynamics on the parameters of the problem and obtain the ionization threshold field strengths as functions of the field frequencies and ionization probabilities for the given field strengths and frequencies.

In such a way we have analyzed the threshold ionization field $F_0^{th}$ dependence on the initial relative frequency $s_0 = \omega n_0^3$ for different values of the quantities $F_{2,0}$ and $s_{2,0} = \omega_2 n_0^3$. As an illustrative example, in fig. 1 we show the results of the calculations for $F_{2,0} = 0$ (one-frequency ionization) and $F_{2,0} = 0.01$, $s_{2,0} = 0.986$ (two-frequency ionization). The parameters of the later case are close to that of the experiment in [7]. However, as far as the purpose of this paper is derivation of the mapping equations for the multifrequency field and analysis of the ionization threshold field dependence on the frequencies but not an analysis of the definite experiments we calculate the "absolute" threshold fields and ionization probabilities (for the infinitely long action of the microwave field with the constant amplitude), i.e. the values do not depending on the concrete experimental conditions: increase and decrease of the field amplitudes, finite time of the fields action, cutoff of the high principle quantum number and so no. Therefore, the parameters in our illustrations are chosen according with the purpose of demonstration the most pronounced resonance structure in the field-atom interaction. So, if, as in ref. [7], we choose $F_{2,0} = 0.0169$ instead of $F_{2,0} = 0.01$ we would observe in fig. 2 a certain (with probability 1) "absolute" ionization probability in the relatively large interval of frequency $s_0$ near the first peak at $s_0 \simeq 0.8$. It should be noted that in paper [7] theoretical results obtained using generalized in some way mapping equations of paper [4] are presented too. The used equations, however, have not been published. Note also that authors of the paper [7] analyze only region of the first peak near the main resonance.

Let us analyze the results of the calculations. So, two curves in fig. 1 corresponding to the one-frequency ionization and two-frequency ionization, (a) and (b) respectively, have similar minimum-maximum structure. The main effect of the second field manifests itself in the shifting of the minimum (maximum) positions to the lower relative frequencies. Besides, due to the influence of the additional field, the first threshold field $F_0^{th}$ minimum (near $s_0 = 0.8$) becomes much more pronounced. The same occurs with the second minimum (near $s_0 = 1.7$), too. The third minimum (near $s_0 = 2.7$) has been, however, almost lost after the supplement of the second field. Note that curve (b) only slightly differs from the curve (c) calculated using correlated initial phases, $\vartheta_0 = \vartheta_{2,0}$. This means that the initial phase difference value is not very important for the qualitative threshold field $F_0^{th}$ behavior. The similar effect has been observed also in the experiment and calculations of ref. [7].

By analogy with the paper [4] we can try to make some analytical evaluation of the threshold field $F_0^{th}$ for the two-frequency case as well. Using the chaos criterion proposed in [9]

$$\max \left| \frac{\delta \vartheta_{j+1}}{\delta \vartheta_j} - 1 \right| \geq 1 \qquad (7)$$

from eq. (5) one gets

$$F_0^{th} = \frac{s_0^{4/3}}{12\pi^2 s_j^{7/3} \mathbf{J}'_{s_j}(s_j)} - \frac{\mathbf{J}'_{s_{2,j}}(s_{2,j}) s_{2,j}}{\mathbf{J}'_{s_0}(s_0) s_0} F_{2,0}. \qquad (8)$$

For $s_j = s_0$ this expression simplifies to the form

$$F_0^{th} = \frac{1}{12\pi^2 s_0 \mathbf{J}'_{s_0}(s_0)} - \frac{\mathbf{J}'_{s_{2,0}}(s_{2,0}) s_{2,0}}{\mathbf{J}'_{s_0}(s_0) s_0} F_{2,0}. \tag{9}$$

The dependences of the appropriate threshold field $F_0^{th}$ on $s_0$ are shown in fig. 1 as (d) and (e) curves for one- and two-frequency field, respectively. For the low initial relative frequency ($s_0 < 1$) eq. (9) is a rough approximation. For relatively high frequencies ($s_0 > 1$) it gives, however, quite accurate threshold field values. The analytical curves are monotonic and, therefore, they represent only approximate threshold field, $F_0^{th}$, behavior and do not reproduce minimum-maximum sequences. For the more accurate evaluation of the threshold field strength according to the criterion (7), the increase of the electrons's energy by the influence of the electromagnetic field should be taken into account [6]. If the scaled energy $\varepsilon_j$ decreases in a result of the relatively regular dynamics in the not sufficiently strong microwave field, then it is enough the lower field strength for transition to the chaotic dynamics.

There is one-to-one correspondence between the threshold ionization field $F_0^{th}$ and the ionization probabilities $P_{ion.}$. Both quantities are measurable. Often it is, however, easier to carry out measurements of the ionization probabilities. Using map (5) we have calculated ionization probabilities for one- and two-frequency fields. Curve (a) in fig. 2 shows one-frequency ionization probability as a function of the initial relative frequency $s_0$ for $F_0 = 0.0298$. Curves (b) and (c) represent two-frequency ionization probabilities for $F_{2,0} = 0.0298$, $s_{2,0} = 0.986$ and $F_{2,0} = 0.01$, $s_{2,0} = 0.18$, respectively, when initial phases of the field components are random. The first peak (near $s_0 = 0.75$) for the two-frequency field with similar to case (b) parameters had been observed in the experiment by Haffmans *et al.* [7]. We see that the second field broadens and shifts ionization probability peaks. Actually, one can realize their behavior from the threshold field curves (fig. 1), too.

We have performed similar calculations for different second initial relative frequency $s_{2,0}$ values. The results show that the first ionization peak is very stable with respect to the relative frequency $s_{2,0}$ changes. This is not the case for the high frequency peaks – their positions and shapes are quite sensitive to the frequency but not to the field intensity variations. To illustrate this, in fig. 3 we represent the ionization probability projection to the frequency $s_0$ and $s_{2,0}$ plane. White color corresponds to the ionization peaks, dark indicates to the ionization valleys. The pattern is almost symmetric with respect to the line $s_0 = s_{2,0}$ although the field amplitudes $F_0$ and $F_{2,0}$ differ considerably. This means that minimum-maximum structure is weakly sensitive to field strengths.

Finally some remarks should be made concerning validity conditions of the analysis based on the map (5). Earlier calculations and comparisons with experimental results [1]-[8], [10] have shown that the one-dimensional hydrogen atom model reproduces quite reasonably the threshold field strength for the one-frequency field in low relative frequency region ($s_0 \leq 1$). The quantum analysis has suggested and experimental results have confirmed that for the high relative frequency of the monochromatic field the classical chaotic diffusion is suppressed by quantum effects [1]-[3], [5]. It is, however, natural to expect that an additional microwave field, like any other external perturbation [11], [12], should result in delocalization of the states superposition and restoration of the chaotic dynamics. Therefore, mapping equations of motion (5) for the multifrequency field should have larger region of validity than those for the monochromatic field.

In summary, using the condition that the phases of the field components are all the time interrelated and in the derivation of the Kepler map we integrate within the period of the electron intrinsic motion, we can easy generalize the Kepler map for the multifrequency field. The dimension of the map does not increase with increase of the number of the field components and, therefore, the analysis based on the maps of the dynamical chaos and ionization by the multicomponent field remains relatively simple. As an example, we have analyzed the map (5) for the two-frequency field and have obtained the threshold ionization field and ionization probability curves (fig. 1 and 2, respectively) for broad relative frequency range, $0.05 \leq s_0 \leq 3$. They show that position of the first peak (near the main resonance) in the ionization probability and corresponding threshold field minimum are weakly sensitive to the second field strength and frequency values. In the high frequency region peaks of the ionization probability are, on the contrary, frequency dependent. From the other side, the minimum-maximum structure is weakly sensitive to field strength values.

***

The research described in this publication was made possible in part by the support of the Alexander von Humboldt Foundation.

**Captions to the figures of the paper**

B. Kaulakys, D. Grauzhinis and G. Vilutis

"Modelling by maps of two-frequency microwave ionization of hydrogen atoms"

Fig. 1. – Threshold field strength $F_0^{th}$ as a function of the initial relative frequency $s_0$. Numerical results are shown: (a) for $F_{2,0} = 0$, (b) and (c) for $F_{2,0} = 0.01$, $s_{2,0} = 0.986$ with random and with correlated initial phases, $\vartheta_0 = \vartheta_{2,0}$, respectively. Curves (d) and (e) represent analytical estimations according to approximation (9) for one- and two-frequency field with the above parameters, respectively.

Fig. 2. – Ionization probability $P_{ion}$ as a function of the relative frequency $s_0$ for: (a) $F_{2,0} = 0$, (b) $F_{2,0} = 0.01$, $s_{2,0} = 0.986$, (c) $F_{2,0} = 0.01$, $s_{2,0} = 0.18$, respectively. In all three cases calculations are fulfilled with $F_{1,0} = 0.0298$ and random initial phases.

Fig. 3. – Ionization probability projection to the plane of initial frequencies $s_0$ and $s_{2,0}$ for $F_0 = 0.0298$, $F_{2,0} = 0.01$ with random initial field phases. Lightness of the background correlates with the ionization intensity: black – no ionization, white – a certain (with probability 1) ionization.

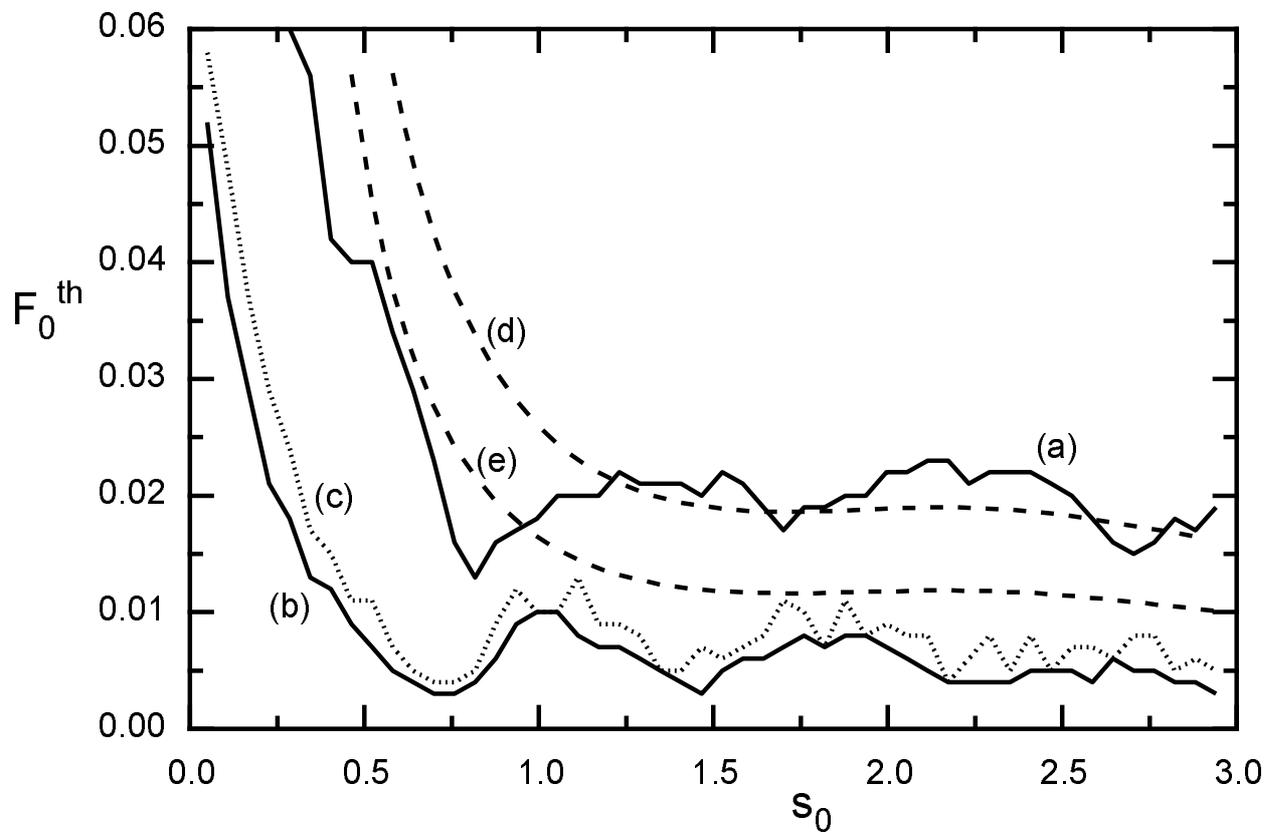

Fig. 1

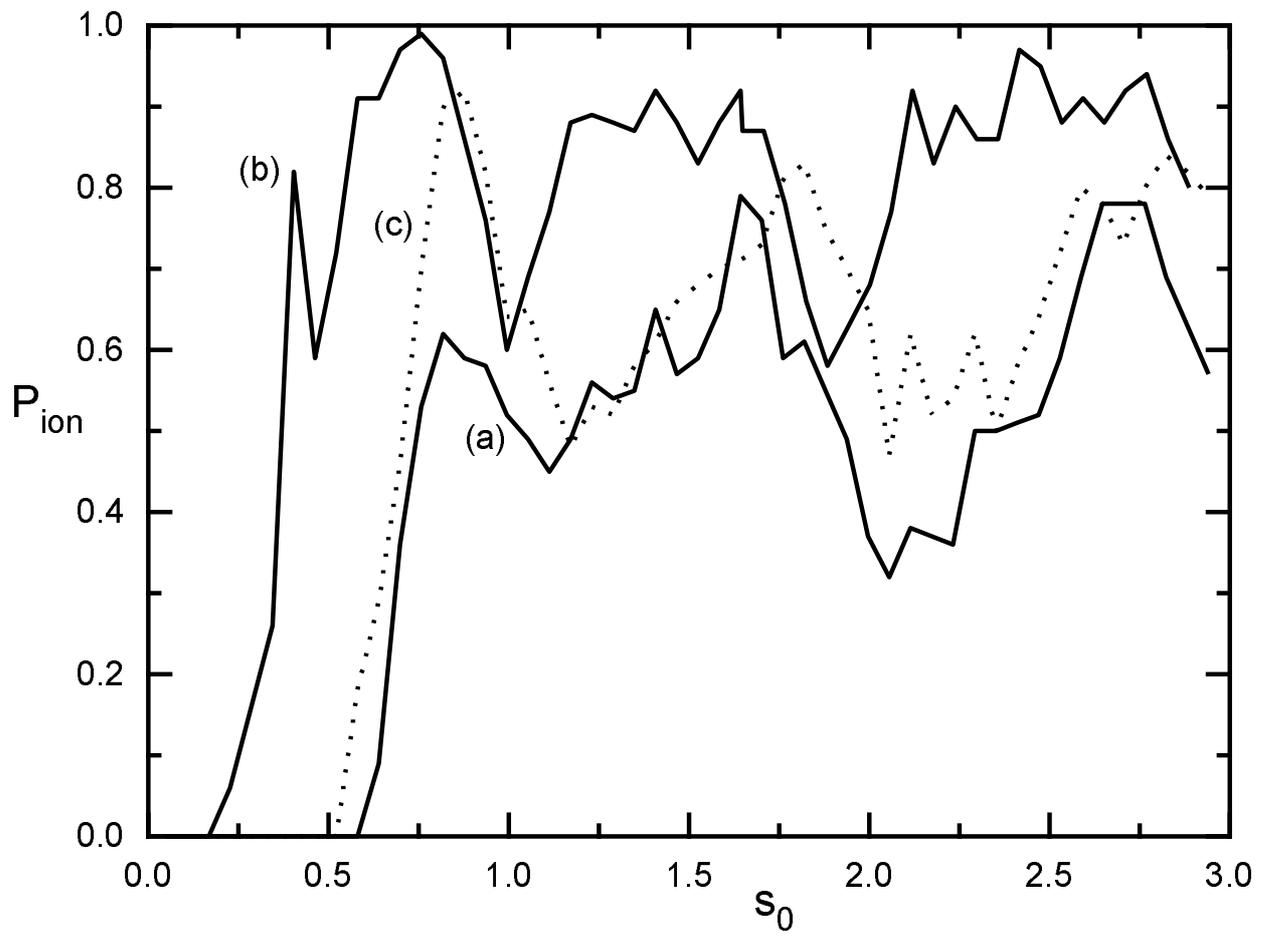

Fig. 2